%% file: document.tex
\documentclass{sig-alternate}

\usepackage{tabularx}
\usepackage{hhline}
\newcommand{\noun}[1]{\textsc{#1}}

\begin{document}

\clubpenalty=10000
\widowpenalty = 10000

\CopyrightYear{2016} 
\setcopyright{acmlicensed}
\conferenceinfo{SafeConfig'16,}{October 24 2016, Vienna, Austria}
\isbn{978-1-4503-4566-8/16/10}\acmPrice{\$15.00}
\doi{http://dx.doi.org/10.1145/2994475.2994479}

\title{An Iterative and Toolchain-Based Approach to Automate Scanning and Mapping Computer
Networks}
%
%
%
%
%

\numberofauthors{3} 
%
\author{
\alignauthor Stefan Marksteiner\\
       \affaddr{DIGITAL - Institute for Information}\\ 
       \affaddr{and Communication Technologies}\\
       \affaddr{JOANNEUM RESEARCH}\\
       \affaddr{Graz, Austria}\\
       \email{stefan.marksteiner\linebreak[0]@joanneum.at}\\
\alignauthor Harald Lernbei{\ss}\\
       \affaddr{DIGITAL - Institute for Information}\\ 
       \affaddr{and Communication Technologies}\\
       \affaddr{JOANNEUM RESEARCH}\\
       \affaddr{Graz, Austria}\\
       \email{harald.lernbeiss\linebreak[0]@joanneum.at}\\
\alignauthor Bernhard Jandl-Scherf\\
       \affaddr{DIGITAL - Institute for Information}\\ 
       \affaddr{and Communication Technologies}\\
       \affaddr{JOANNEUM RESEARCH}\\
       \affaddr{Graz, Austria}\\
       \email{bernhard.jandl-scherf@joanneum.at}\\       
}

\maketitle
\begin{abstract}
As today's organizational computer networks are ever evolving and becoming more and more complex, finding potential
vulnerabilities and conducting security audits has become a crucial element in securing these networks. 
The first step in
auditing a network is reconnaissance by mapping it to get a comprehensive overview over its structure.
The growing
complexity, however, makes this task increasingly effortful, even more as mapping 
(instead of plain scanning), presently, still involves a lot of manual work.
Therefore, the concept proposed in this paper automates the scanning and mapping of unknown and non-cooperative 
computer networks in order to find security weaknesses or verify access controls.
It further helps to conduct audits by allowing comparing documented with actual networks and finding
unauthorized network devices, as well as evaluating access control methods by conducting delta scans.
It uses a novel approach of augmenting data from iteratively chained existing scanning tools 
with context, using genuine analytics modules
to allow assessing a network's topology instead of just generating a list of scanned devices.
It further 
contains a visualization model that 
provides a clear, lucid topology map and a special graph 
for comparative analysis. 
The goal is to provide maximum insight with a minimum of a priori knowledge.
\end{abstract}

%
%


%
%

%
%



\input{content}

\section{Acknowledgements}
\label{sec:ack}
This research has been conducted on behalf of the Austrian Federal Ministry of Defense and Sports (BMLVS), Joi-nt
Command Support Centre (F\"{u}UZ), section ICT technology, department ICT security and funded by the Department for
Science, Research and Development (WFE). We want to thank all partners, especially Lambert Scharwitzl, Christina Buttinger,
Michael Pfister and Clemens Edlinger for their support.

\bibliographystyle{abbrv}

\end{document}

%% file: content.tex
\section{Introduction and Motivation}
Scanning and mapping a computer network is the crucial part of network reconnaissance \cite{convery2000cisco}.
As such, it is usually one of the first steps performed by an attacker\cite{MR:2004}. 
Therefore, it is also useful for auditors or people in charge of network security to use the same techniques to
examine their network in order to identify potential security weaknesses \cite{muelder2005interactive}. 
This overview differentiates in this context also network scanning and mapping. While the former generates solely a
list of network devices, the latter generates a map, providing additional information about the network's structure,
building the fundament to measuring security in the form of attack graphs \cite{4041160}. 
To support this mapping task, this paper from an ongoing research project presents a widely automated
approach for mapping non-cooperative, previously unknown computer networks and an expressive visualization of the
obtained data.
Within this
paper, \textit{unknown} means that there is only minimal a priori knowledge of the network, in particular the target
network range to be scanned (or even only parts of it, like a single machine's address, from that range).
\textit{Non-cooperative} in this context means that the scanner does not have any administrative access (such as
credentials).

Similar to testing applications with unknown internals, these conditions can also be subsumed as \textit{black box
scanning} \cite{shelly2010using,bau2010state}.
The benefit of black box scanning non-cooperative networks in security auditing and ethical hacking is threefold:
firstly, it allows comparing the network documentation with the actual reality in a company's network. Secondly,
undocumented, unauthorized and possibly malicious devices may be discovered \cite{orebaugh2011nmap}. And thirdly, as
mentioned above, it provides an attacker's view \cite{BDA:2014}.
All three of the above points (but especially the last one) benefit from an approach with only minimal a
priori information (black box), as it prevents the security staff from taking paved roads in their analysis and, thus,
from gaining the same results they already obtained beforehand by more white box oriented security evaluation forms.
Also, network topology is among the needed information for creating an attack graph
\cite{SP:1998}.
There is a broad variety of network and port scanners available (for instance nmap\footnote{\url{www.nmap.org}} or
amap\footnote{\url{www.thc.org/thc-amap}}), as well as specific vulnerability scanners (like Nessus\footnote{\url{www.tenable.com/products/nessus-vulnerability-scanner}} or OpenVAS\footnote{\url{www.openvas.org}}) or
tools for further information gathering (as for example
snmpwalk\footnote{\url{www.net-snmp.org/docs/man/snmpwalk.html}}).
Unfortunately, the process of performing a thorough network reconnaissance (that results in a topology map) is,
presently, rather manual and time-consuming.

Therefore, this paper addresses the problem of how aforementioned network scanning tools can be most effectively
combined and their scanning results analyzed and graphically represented.
In order to fulfill this objective, the following questions need to be clarified:
\begin{itemize}
\item How can the scanning process be automated and controlled?
\item How can the results be automatically analyzed?
\item How can the results be best graphically represented in a scalable and lucid form?
\end{itemize}
The third question has the additional constraint that large networks should be presented in an aggregated manner, but
still leave the basic topology visible at a glance
, which existing solutions (see Section \ref{sec:rw}) have proven not to meet.

Also, as one of the most relevant aspects in analyzing computer networks is
discovering and visualizing the network's structure, the second question contains the major problem of augmenting the
obtained data with structural information.
Modern computer networks consist of nodes (hosts and routers) and edges (links) that 
 are structured in a hierarchy, which means that their structure
can be represented as a tree (that is, as a directed acyclic graph
with a single root) as defined by graph theory \cite{RFC4632}. 
Therefore, the pivotal element of determining a node's place within that structure is
finding the edge to its parent node or, in computer networking terms, its default gateway. This leads to the
following subquestion that defines the step from list-like network scanning to actual network mapping by adding hierarchical
context to the scans:
\begin{itemize}
  \item How can a node's default gateway be determined?
\end{itemize} 

This problem statement applies, in principle, to all computer networks. However, the presented solution is shaped
to fit to modern day's organizational (in other words: classical) information and communication technology (ICT) networks.
Specific requirements for other types of networks (for instance within industrial control systems or the Internet of
things) are out of scope of this work. A main concept used to automatize the scanning process is chaining these popular
network scanning tools (see Section \ref{sec:tc}).
Tool chaining is a well-known concept in information technology and also (although less commonly) used in 
approaches to network scanning in order to yield better results than there could be gained from a single scanning tool
(see Section \ref{sec:rw}).
The presented solution further develops this technique by
enhancing the versatility of the scanning process introducing \textit{scanning policies}. These policies contain
customizable toolchains improved with an iterative scanning model, allowing each
pass of the said toolchain to benefit from the results obtained by its predecessor (see Section \ref{sec:it}). 
Although such a concept is used in white box network mapping (often called \textit{network discovery} in this context),
the approach is quite novel to black box network mapping (see Section \ref{sec:rw}).
Scanning policies further integrate external scanning tools and genuine analytics to automatically interpret the
scanning results, yielding a mostly automatically generated topology model (see Section \ref{sec:tc}).
Furthermore, as the analysis and subsequent graphical
representation still are less well elaborated parts in mapping non-cooperative computer networks in general, this
approach also includes a visualization concept for the data obtained through scanning and the contextual enhancements
generated by the analytics modules (see Section \ref{sec:vis}). The visualization concept includes automatic grouping of
the data (grouped by operating systems or network segments), while leaving network infrastructure nodes, and therefore
the overall topology, visible.
It also features a comparative view, allowing documenting network evolvement over time and testing of access rules.
Apart from an approach to solving the questions stated above, this work also provides a proof of concept in the form of
a software prototype of the presented concept (see Section \ref{sec:sa}) and scan results of
 real-world  networks (Section \ref{sec:rs}).
Section \ref{sec:cn}, eventually, gives a conclusion and outlook to further research.

\section{Related Work}
\label{sec:rw}
There is a known approach to toolchaining in black box scanning that has some similarities in
terms of using adapter plugins for scanners and splitting normalization and aggregation \cite{CRS:2011}.
That work focuses on building an integrated scanning tool and merging the obtained
information in a common structure, while the presented research, in contrast, does not merge the information but
displays each scan result distinctly in order to distinguish which module found which information. The main difference,
however, is that this paper's approach focuses on network mapping (including scanner analytics and visualization)
instead of the scanning part and uses an iterative model. The visualization concept in this paper is the result of an evaluation (see
Section \ref{sec:vis}) of a variety of commonly known  algorithms
\cite{ward2015interactive,Johnson1991,Reingold1981,JankunKelly2003,Lin2006,Carriere1995,grivet2006bubbletree}, as
commonly used graphical scanner interfaces (such as \textit{Zenmap}\footnote{\url{www.nmap.org/zenmap}}) have proven to be insufficient for network mapping.
Using an iterative model, on the other hand, is common in commercial and open source solutions for credential-based
white box network mapping\footnote{often called network discovery in this context, particularly in solutions for network
management}, but rarely used in black-box scanning. In fact, only one approach came to the authors' attention
\cite{boyter2003system}, which again does merely scan but not map unknown networks.

\section{Toolchains}
\label{sec:tc}
This section describes a concept for tool chaining in the context of computer network scanning and mapping.
In this approach, the building blocks of such toolchains are, in principle, scanner modules and analyzer modules. The
two differ mostly in two points: their purpose and their input method. Firstly, while scanner modules represent existing
network scanning tools, the purpose of analyzer modules is to provide context for the scanned data. Secondly, scanner
modules get external input from their respective scanning tools, while analyzer modules merely operate on the
present data without external input.
\begin{figure}
	\centering
		\scalebox{0.4}{
			\includegraphics{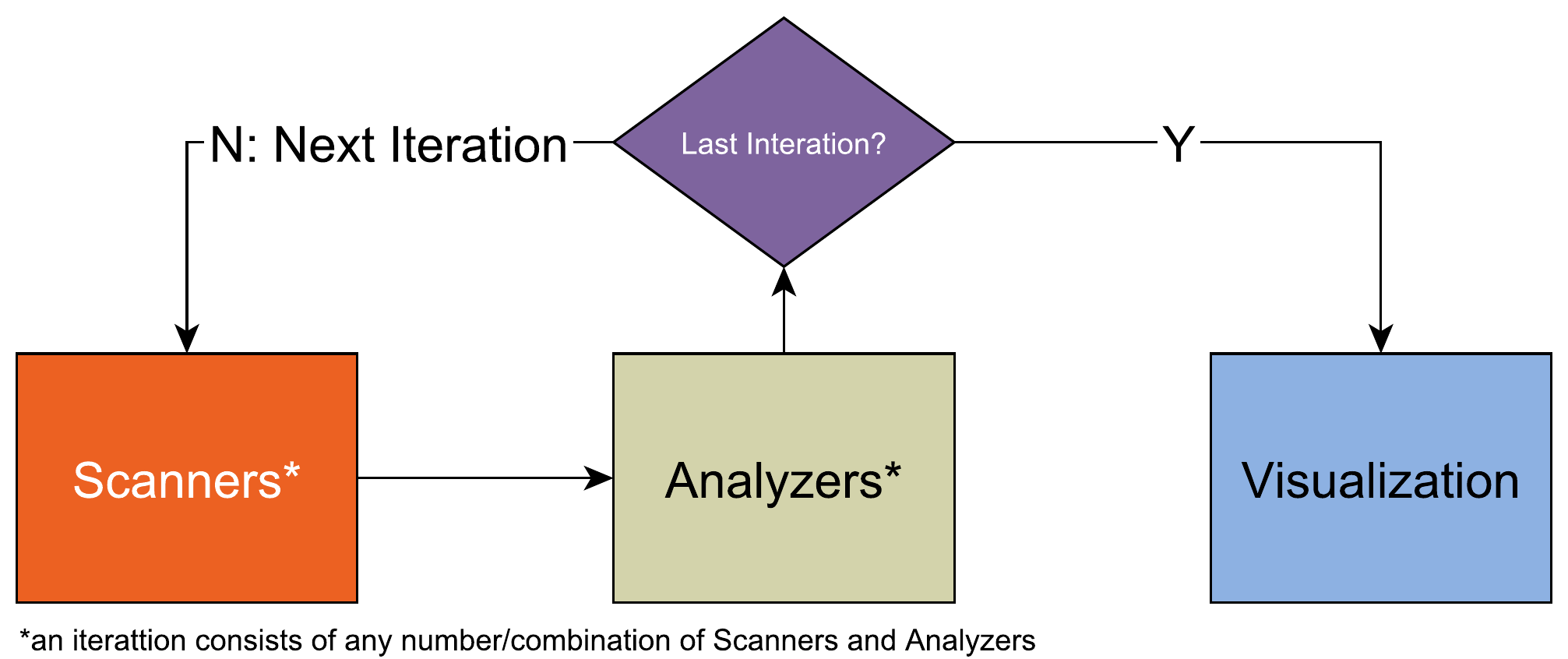}
		}
	\caption{Schematic example toolchain with iterations}\label{img:it}
\end{figure}
As seen in Figure \ref{img:it}, 
these two elements are the building blocks of 
a \textit{scanning policy} which, within this approach, represents a toolchain and the number of
iterations the chain is passed through (see Section \ref{sec:it}). 
Using policies, toolchains 
are not static and can be adjusted according to specific needs, as different toolchains might yield different
results (even the same tools used in a different order). The same applies to starting options for both scanner and
analyzer modules. This flexibility allows choosing the most effective policy for each case (for instance,
different chains for small and big networks).

\subsection{Scanner Modules}
\label{sec:tc:scan}
\begin{figure}
	\centering%
		\scalebox{0.48}{
			\includegraphics{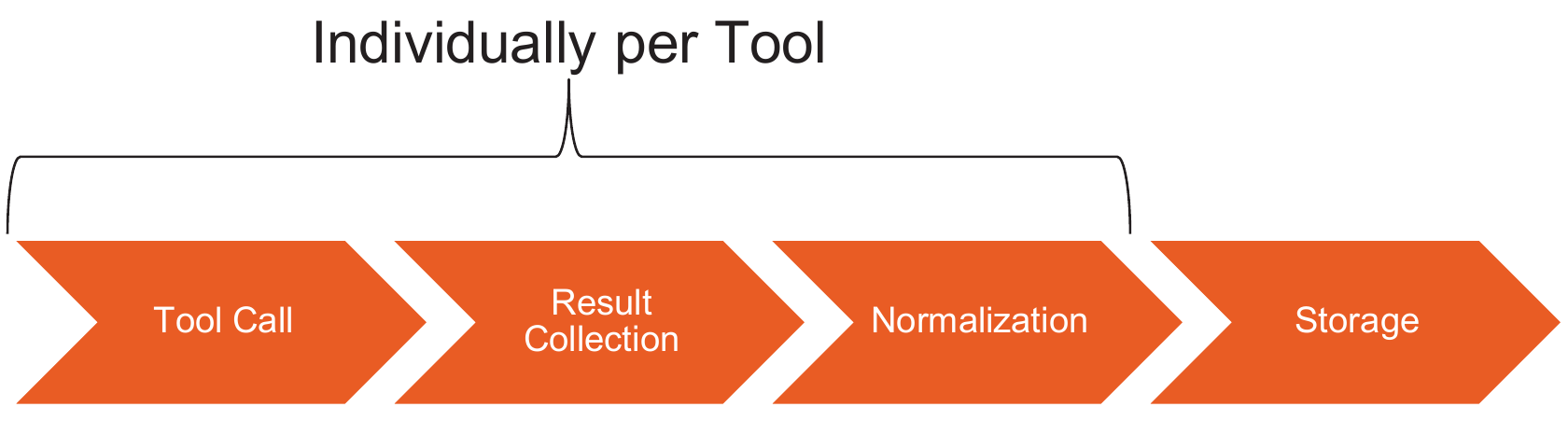}
		}
	\caption{Execution trail of a scanner module}\label{img:mod}
\end{figure}

A scanner module controls an external network scanner. Its tasks are therefore handling the scanning tool's input
(tool call) and output (result collection), converting the output into a common format (normalization) and storing it in
a database (storage). This process is the same for every module; only the details differ (except for the storage, see
Figure \ref{img:mod}).
Calling a tool could happen either through command line calls, an API or by incorporating the tool's source code
directly into the module's. Either way, a module is responsible for enabling all sensible (in the context of a
toolchain) options its tool provides, for defining additional ones (mostly optimizations for better interlocking of tools within a
toolchain) and for using some options per default (for example output formatting options). Consequentially, the next step
is collecting the results delivered by the scanner software, either through a return value (if called by code or API) or
(if otherwise) by reading a file or console output. These results then must be normalized to correspond to each
other. As it provides a broad range of already present attributes, this solution uses the
\textit{nmap xml output} as a basis \cite{lyon2008nmap}. Attributes not present in this format are incorporated as
needed by other modules. For storage reasons stated in Section \ref{sec:sa}, the normalized data is subsequently
transferred into the \textit{JavaScript Object Notation (JSON)} \cite{RFC7159}. The combination of obtained data consists of
two components:
normalization (see above) and aggregation (linking the different modules' results together). 
In this approach, the aggregation is a
matter of the visualization engine. The data model creates a separate object per tool for each found node in order to make it clearly
distinguishable which scanning tool yielded which results. This is the reason why scanner modules only handle the
normalization and not the aggregation. In general, newer results take precedence over older ones (except for manual
edits). The storage, eventually, basically consists of writing the obtained results into a
database.

\subsection{Analyzer Modules}
\label{sec:tc:ana}
In contrast to scanner modules, analyzer modules do not use external data, but only information provided by scanner
modules. Therefore, a toolchain containing only analyzer modules will not produce any results. The purpose of analyzer
modules is interconnecting information and providing context. These modules generate database entries containing
the analysis results, which again can be used by subsequent modules, in later iterations or by the visualization engine.  
The main reason not to incorporate the analytical logic within scan modules is that data from several scanners might be
used as input for a specific analysis. For instance, network trace information might come from an arbitrary
\textit{traceroute} implementation, as well as an \textit{nmap} (or other trace-capable) scanner module.  
The following subsection gives an example for an analyzer module.

\subsection{Estimating Default Gateways}
\label{sec:tc:ana:sg}
Estimating default gateways is the necessary stepping sto-ne to overcome the gap between network scanning and network
mapping. As the required data to achieve this goal might be present within the acquired data but has to be interpreted
correctly, this task poses an archetype for an analyzer module in the sense of the presented concept.
To fulfill this task, this work proposes three methods to determine a host's default gateway $DGW(H)$:
\begin{itemize}
  \item \textit{Estimation by trace}: Determining predecessor information in a network trace\cite{957902};
  \item \textit{Estimation by singleton router}: Assessing if only one router is present in a scanned network;
  \item \textit{Estimation by usual suspects}: Watching for addresses frequently used for gateways.
\end{itemize} 
The first method naturally requires a list of hops ($h$) obtained through traceroute from a scanner in the
toolchain.
If the host is in the same subnet (hopcount $n=1$), the host will be attached to the scanning host's own default gateway
$DGW(X)$, which is determined using system functions.
Otherwise, the default router of the host is set to the last hop before the host ($h_{i-1}$, see Formula \ref{eq:dgw}).
\begin{equation}\label{eq:dgw}
i:\{1,..,n\}; DGW(H) = \left\{ \begin{array}{ccl}
DGW(X) & \ldots&n = 1\\ 
h_{(i-1)} & \ldots&n > 1\end{array}\right.
\end{equation}
As the path towards the target network holds no information about its topology and is therefore of less interest, an
algorithm identifies the last common gateway of the network. This \textit{network entry point} (from the scanner's
perspective) of all hosts $NEP(H)$ serves as a central connection node that links different parts of a graph
together (for an example, see Section \ref{sec:rs}). Said algorithm traverses all of the trace paths $h_i(H_j)$
in parallel from their beginning and compares if they are all equal at the respective hop position. The last common
router is the router at one position ($h_{i-1}$) before they begin to diverge, which means that at least one path
differs at a hop position ($h_i(H_j)\neq h_i(H_{j+1}$).
Therefore, if not all paths contain the same address value at hop position x, the last common
gateway is x-1. If hosts from the scanner's network are present in the target (with $n=1$),
the result is the scanning machine's default gateway (see Formula \ref{eq:nep}). 
\begin{equation}\label{eq:nep}
\begin{split}
i:\{1,..,n\},j:\{1,..,m\};\\ 
NEP(H)=h_{i-1} \leftrightarrow h_i(H_j) \neq h_i(H_{j+1})
\end{split}
\end{equation}
The second method yields a positive result if exactly one device of some \textit{router} type is present in the
dataset. This requires operating system (OS) detection and classification capabilities by a scanner in the toolchain.
For instance, nmap provides this feature in its \textit{nmap-os-db}. The analyzer module examines this data for
router devices.
The third method is similar to the second one in terms that it also needs the same information to be present and
searches for \textit{router}-type devices. In contrast to the latter, it uses the first three \textit{dotted decimal}
groups of IPv4 target addresses and adds commonly used router addresses (e.g. \textsl{.1} and \textsl{.254}) as final
part.
These composites form the addresses that are evaluated to be a router. Further, for multiple targets, duplicates are eliminated. If such a \textit{usual gateway address} is
in fact a router according to OS detection, there is a distinct probability of being a default gateway for the scanned
network.

The first method is more reliable (as a traceroute predecessor is more likely a default gateway router than an
arbitrary one found on the same network), but the success of these methods obviously depends on the present information
and, thus, ultimately on the target network's configuration. As the latter is unknown beforehand, it cannot be known a
priori which method is more successful. As none of the estimation methods provides assured detection, the opportunity
for manual assignment and correction (through user interaction, superseding automated scan results) is included in this
concept.

\subsection{Iterative Scanning}
\label{sec:it}
The basic idea described in this section is to let the results of one
iteration determine the input for the next one \cite{boyter2003system}. To do so, every module within the
toolchain implements its own method for gaining \textit{seed} data to be used in the next iteration. This method can
differ greatly between modules. For instance, one tool might use network trace data, while another scanning
module might use data gained via \textit{SNMP} or \textit{ARP} (Section \ref{sec:rs} shows a practical example). 
The default gateway estimation module (see Section \ref{sec:tc:ana:sg}), for instance, adds identified intermediate hops
that are not yet scanned to the seed database. This usually yields an expanded scan area, for intermediate hops
frequently feature IP addresses from outside the target scope.
Subsequently, all newly found hosts will serve as targets for the next scan iteration, thus being processed by
all tools within the chain (see Figure \ref{img:it}). Together with the respective toolchain, the number of iterations
forms a scanning policy.

\section{Network Visualization}
\label{sec:vis} 
Due to the potential of visualizations
\cite{Burkhard2004} and the speed of visual perception \cite{Liu2009281}, it seems appropriate to 
present a computer network in graphical form.
Various solutions for automatically generating a graphical view of
a network hierarchy through a computer program have been proposed (see Section \ref{sec:rw}).
In order to find the appropriate algorithm for computing the graph's
layout the following requirements have to be considered: 
\begin{figure}
\centering\includegraphics[width=2.5in]{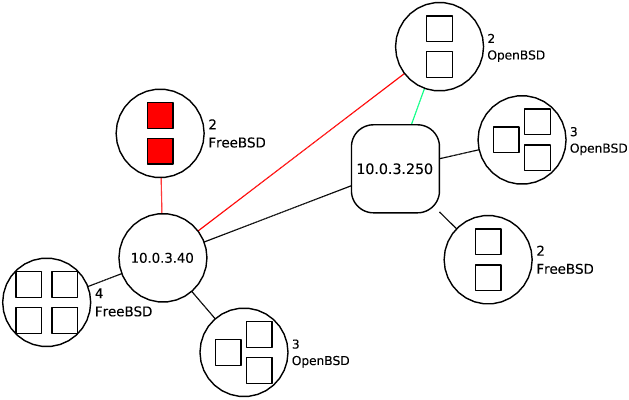}
\caption{Compare view showing gateway change and node removal}
\label{img:comp} 
\end{figure}
\begin{itemize}
\item Suitability for typical organizational network sizes;
\item Economic space usage, even for unbalanced hierarchies; 
\item Edges shall contain as few bends as possible; 
\item Change tolerance regarding the overall layout.
\end{itemize}
Using the Tulip graph analysis program, layout algorithms for hierarchies were evaluated 
on these requirements with data collected from a scan of a part of the Joanneum Research
computer network, as well as with generated and manually edited data. Most promising candidates were: 
\begin{itemize}
\item The Tree Radial algorithm \cite{JankunKelly2003};
\item The Balloon algorithm \cite{Lin2006,Carriere1995}; 
\item The Bubble Tree algorithm \cite{grivet2006bubbletree}.
\end{itemize}
The first algorithm's adequacy depends on the parameters for node spacing and layer spacing, while the second works well
for rooted trees if not enforcing equal angles, but has deficiencies in usage of available space.
The Bubble Tree algorithm shows the best overall results, for the graph is laid out with good usage of screen
space, edges contain at most one bend and the overall layout proved to be quite insusceptible
to minor changes.
 All of these layouts, however, do not set focus on a network's structure as they value the structure-defining
intermediate nodes (routers) the same as the usually vastly outnumbering leaf nodes (hosts). To emphasize the former,
similar leaf nodes are aggregated and depicted only by one representative symbol (bubble) with a single edge. This
aggregation could fund on similar characteristics (like the same OS) or on a threshold of the size of leaf
node groups (see Figure \ref{img:itres} for a practical example). Further, leaf nodes could be hidden completely and
only displayed on demand.
For auditing and evaluating a network, not only a snapshot but also displaying deltas of a network is useful in
order to document its evolvement or evaluate the effectiveness of network access control. To do so, an adapted
version of the network graph view is used, highlighting the changes (red for removals and green for additions, see 
Figure \ref{img:comp} for a concrete example).

\begin{figure}
	\centering%
		\scalebox{0.4}{
			\includegraphics{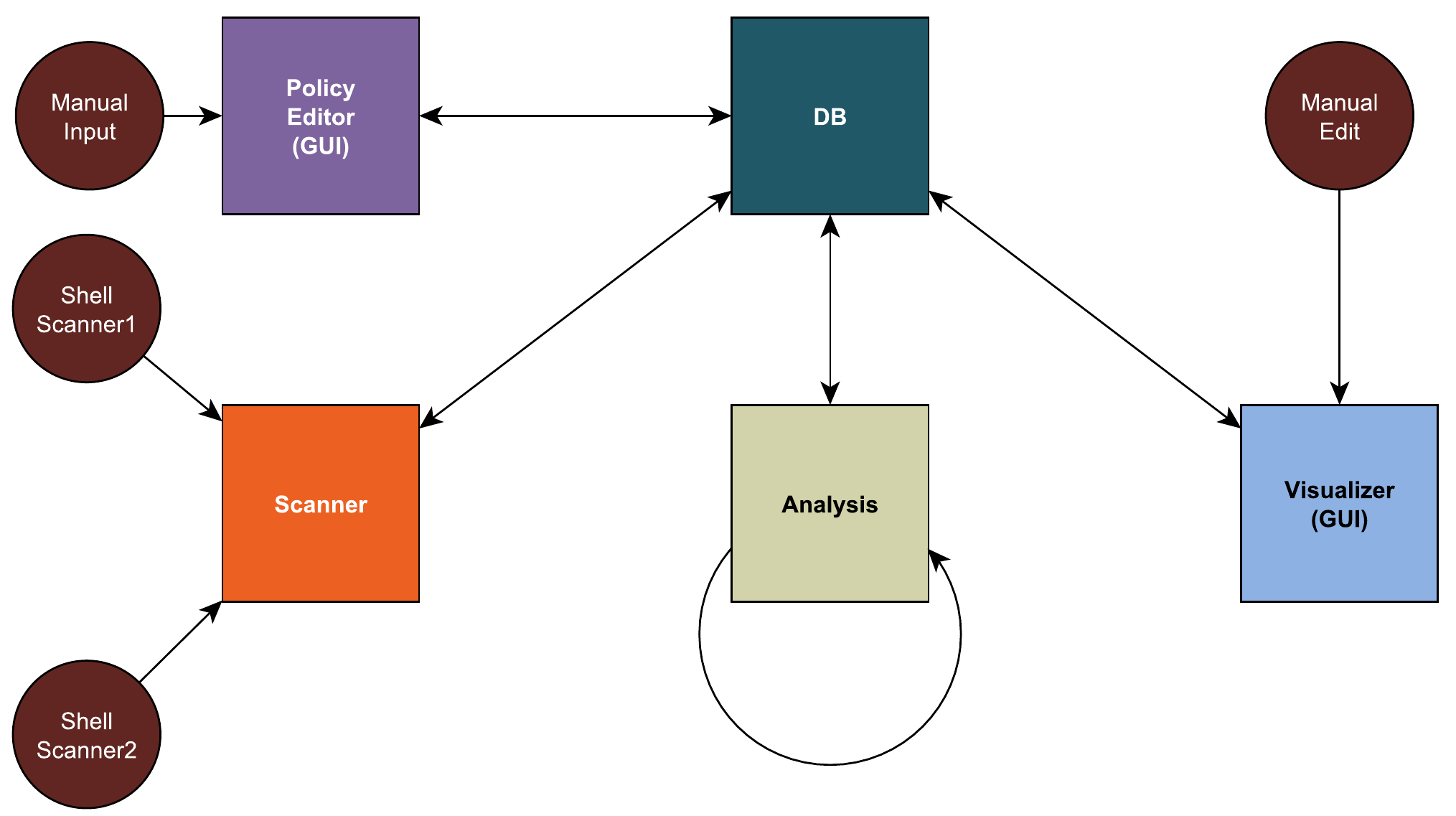}
		}
	\caption{Diagram of the module interaction}\label{img:arch}
\end{figure}
\section{Software Prototype}
\label{sec:sa}

This section describes a software prototype implementing the concepts illustrated above in C++,
resulting in a tool for Kali Linux\footnote{\url{www.kali.org}}.
This software, called \textit{Tactical Network Mapper
(TNM)} uses MongoDB\footnote{\url{www.mongodb.com}} and the Tulip graph library\footnote{\url{tulip.labri.fr/TulipDrupal/}}
for storage and visualization purposes, respectively. 
The main libraries of the software contain only a framework, while the plug-ins provide the functionality, resembling
the structure of the Eclipse\footnote{\url{www.eclipse.org/ide/}} software development environment. 
All the data exchange between the components (scanner and
analyzer modules, and user interface including policy editor and the visualization engine) happens via the database (see
also Figure \ref{img:arch}). This couples these components very loosely, allowing a very flexible plug-in structure. As
a result, the built solution allows integrating rather heterogeneous modules, making it an adequate testbed 
for both the current and future concepts.
\begin{figure}
	\centering%
			\includegraphics[width=3.32in]{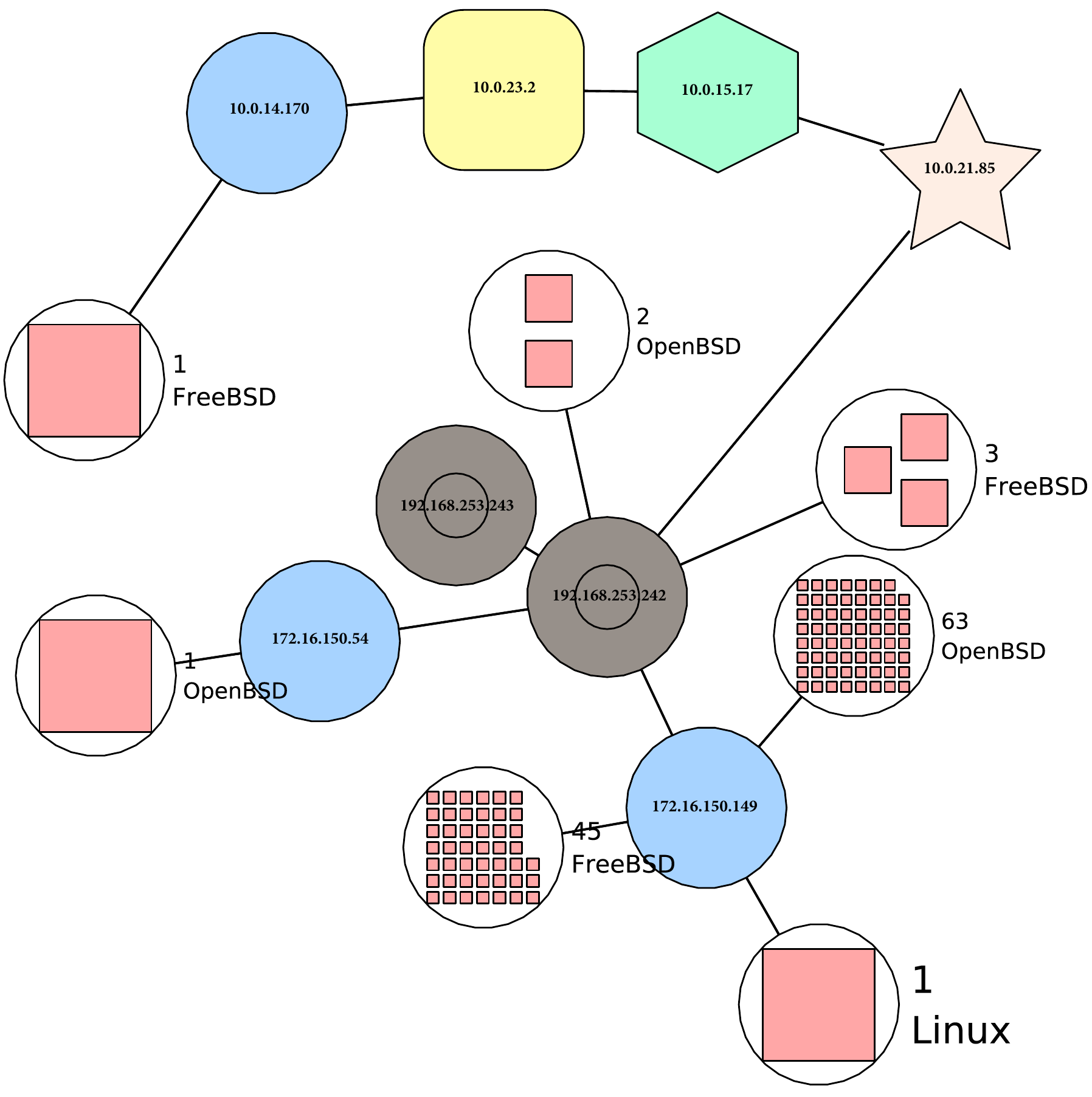}
	\caption{Graph of an anonymized external actual scan}\label{img:ext}
\end{figure}
As the scanning process is iterative and a user might also manually revise the results and make changes, not only
the most recent information might be relevant. 
A user could want to view changes between iterations and possibly roll back to a previous state, as well as to
manually correct data (e.g. in case of wrongly assigned default gateways).
These requirements point towards versioning as used in Revision
Control Systems \cite{ruparelia2010history}.
There are two main approaches to versioning: \textit{snapshot} and
\textit{changeset (delta)} \cite{KT:2011}. The former method stores a complete copy of the data set
for each version. The latter saves storage space, but requires the construction of data sets in order to get a certain
version. The changeset group can be further
classified into \textit{forward delta} and \textit{reverse delta}. With reverse delta, always the latest version serves
as reference for differences \cite{wong2002managing}. Despite earlier versions might be needed for comparing, it is
reasonable to assume that most users will be interested in the most recent version of the data. 
Therefore, the presented solution tries balancing the benefits of both approaches by
using \textit{snapshots} for single objects, but \textit{reverse
delta} storage for the data set as a whole.
This can be easily implemented by storing full snapshots of changed objects (and only those), with the most recent data set as
reference. Through using MongoDB, which uses a binary-encoded \textit{JSON}
format (\textit{BSON}) for storing the data \cite{membrey2011definitive}, no additional conversion of the
normalized data is needed.

\section{Results}
\label{sec:rs}
As the concept (in form of the \noun{TNM} implementation) is put to test, the results expectably differ according to
different target settings.
Obviously, the obtained results are more detailed when scanning from inside a network. This is a practical scenario for
internal audits. Despite some natural restrictions (access rules enforced by firewalls), \noun{TNM} was able to perform
on external networks too. Figure \ref{img:ext} shows an anonymized example of a scan executed on an external, unknown
(that is with only the target range known) network using an \textit{nmap} module with traceroute and OS detection and
the default gateway analyzer.
The scanning machine was attached to a router not shown in the graph (according to the algorithm described in Section \ref{sec:tc:ana:sg}). Through the iterative model, \noun{TNM} also
discovered parts of the network that where originally not part of the target specification, which causes the different
network portions of the IP addresses seen in Figure \ref{img:ext}.

\begin{figure}
	\centering%

			\includegraphics[width=3in]{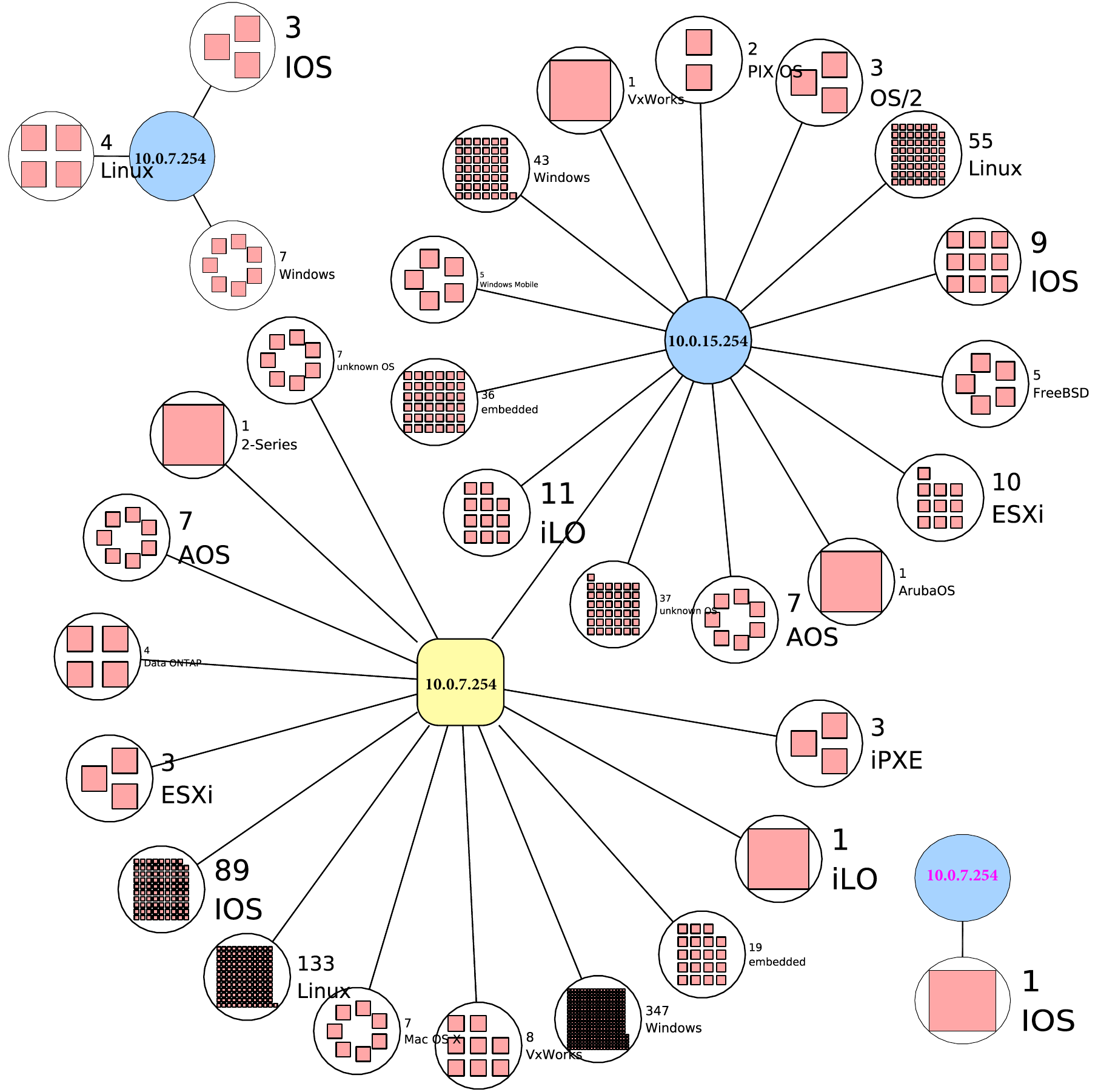}
	\caption{Comparison of three iterations of an anonymized actual scan result graph}\label{img:itres}
\end{figure}
Another example for the use of iterations can be seen in Figure \ref{img:itres}, which shows three iterations of an
internal scan starting with a single machine. The bigger graph in the middle shows the third iteration, while the
adjunct graphs in the bottom right and top left corners show the first and second iterations, respectively. This scan
used a module that implements an snmpwalk scanner. The toolchain contains a TCP \textit{nmap} scan containing traceroute and OS detection,
the default gateway analyzer, a second \textit{nmap}, scanning for the SNMP port (\textsl{UPD/161}) and the snmpwalk
module that is executed on found snmp hosts.
As many products are
shipped with a \textit{public} community, this could be used to discover devices reading out a device' ARP table, which was assumed in
this example \cite{chatzimisios2004security} (simulated by using a known write-only community). 
Clearly visible,
each of the three iterations discovers significantly more devices than the last one (2, 7 and 856 nodes, respectively), eventually resulting in a tree of depth
three.
An
additional scan on a public \textsl{/19} network conducted with TNM (same policy as above) yielded 673 reachable
hosts, 16 open SNMP devices, but no communities could be found.

Naturally, the number of discovered items has an impact on an iteration's runtime.
The first iteration of said test took seven seconds to finish, the second already took forty-four, while the 
last iteration took over half an hour (1996 seconds).
Table \ref{fig:table:Perf} shows an example table for test results (discrepancies between module and total runtimes are
mostly database-induced delays).
It can also be seen that, the ratio of time consumption
and results (found nodes per second) improves significantly at each iteration.
\begin{table}
 	\caption{Runtimes per module and iteration in seconds compared to found nodes}
 	\label{fig:table:Perf}
 	\centering%
	\begin{tabularx}{0.475\textwidth}{|l|X|X|X|}
			\hline
			{Iteration}		& \textbf{1}		&\textbf{2}		&\textbf{3}\\
			\hline
			{NmapScanner(-A)}		&{7}		&{27}		&{1114}\\
			\hline
			{DefaultGatewayAnalyzer}		&{<1}		&{<1}		&{<1}\\
			\hline
			{NmapScanner2 (UDP:161)}		&{<1}		&{2}		&{8}\\
			\hline
			{SnmpwalkScanner}		&{<1}		&{43}		&{797}\\						
			\hline
			{Module}s& {7}&{43}&{1920}\\
			\hline			
			{Total} 		&{7}		&{44}		&{1996}\\
			\hhline{|=|=|=|=|}
			{Number of found nodes}&	{2}&{15}	&{856}\\
			\hline
			{Nodes per second}&	{0.29}&{0.34}	&{0.43}\\
			\hline
		\end{tabularx}
\end{table}
An iteration's scope 
allows a rough estimation of the upper boundary of its runtime, assuming no unforeseen events
(for example network congestion) occur. 
More precise estimations on the runtime of an iteration are not
possible a priori, as it depends on the result (the same applies for the overall process).

\section{Conclusion}
\label{sec:cn}
The presented results demonstrate the successful automated topology mapping of given 
computer networks (both internal and external\footnote{Although there are be limits by filter mechanisms and NAT}), 
without information beyond the target range.
Especially, no credentials are needed (although utilized, for example in SNMP scanning).
They further incorporate additional information about these nodes (operating system, open ports, et cetera) into the
produced interactive map.
This combination means a significant advancement compared to drawing topology maps and maintaining the result lists of
unknown networks manually, using scanning tools and their result lists. 
It is also a novel approach, as no other known
concept combines iterative toolchaining with analytics to conduct black box network mapping and tailored
state-of-the-art visualization.
 By doing so, the presented concept could not only be used in practice to perform security audits and discover
differences between documentation and reality as well as map undocumented networks, but also as a documentation tool
by itself.
The test results show that the concept is capable of not only mapping a network, but also
discovering areas not found using less sophisticated methods. Furthermore, they demonstrate that using more
iterations yields a more effective scanning performance (more found nodes per second of runtime).

Despite promising results, there is room for improvement. 
Apart from additional scanning modules, one possible
upgrade is a module that allows the software to reconfigure the network address of the scanning machine by using data from
passive scanners (or \textit{network sniffers}). This way, the solution is capable of mapping a network it is directly
attached to completely without prior knowledge (eliminating the need for a scan target). Useful for this improvement
(but also for other cases) is a (yet work in progress) module, that tries to guess network boundaries (partially by
using publicly available data like \textit{whois databases}). Another work in progress module is an extension that
fetches data from vulnerability databases, like the \textit{National Vulnerability Database
(NVD)}\footnote{\url{nvd.nist.gov}}, and automatically compares the contained vulnerable services and operating
systems with data from the scan results to facilitate vulnerability analysis. Another module could use graph theoretical
methods to gain knowledge about the network's structure. Further, the concept could be tested in and adjusted to other
environments like industrial control and Internet of things networks.

Finally, the resulting TNM software itself could also serve as a research tool.
Possible studies carried out using \noun{TNM} include the evaluation of the performance of different combinations of scanning tools
under different conditions (for example with different network sizes).
A prerequisite for such studies is the definition of sensible \textit{key performance indicators (KPIs)} for network scanning and using respective toolchains under different
conditions. Also, surveys on networks suggest themselves, like typical structures depending on network sizes or the rate
of open snmp communities in average structures.